\documentclass[prd,preprint,amsmath,amssymb,superscriptaddress,floatfix,nofootinbib,11pt]{revtex4-1}
\usepackage[utf8]{inputenc}
\usepackage{mathrsfs,bm}
\usepackage{natbib}
\usepackage{txfonts}
\usepackage{amssymb}
\usepackage{rotating}
\usepackage{indentfirst}
\usepackage{graphicx,booktabs}
\usepackage{multirow}
\usepackage{slashed}
\usepackage{overpic}
\usepackage{color}
\usepackage{amssymb}
\usepackage{hyperref}
\usepackage{bbding}
\usepackage{url}
\begin{document}
\title{Magnetic moments of the spin-$\frac{3}{2}$ doubly charmed baryons in covariant baryon chiral perturbation theory}

\author{Rui-Xiang Shi}
\affiliation{School of Physics,  Beihang University, Beijing 102206, China}

\author{Li-Sheng Geng}
\email[E-mail: ]{lisheng.geng@buaa.edu.cn}
\affiliation{School of
Physics  \& Beijing Key
Laboratory of Advanced Nuclear Materials and Physics,  Beihang
University, Beijing 102206, China}
\affiliation{School of Physics and Microelectronics,Zhengzhou University, Zhengzhou, Henan 450001, China}

\begin{abstract}

Inspired by the  discovery of the spin-$\frac{1}{2}$ doubly charmed baryon $\Xi_{cc}^{++}$ and the subsequent theoretical studies of its magnetic moments, we study the magnetic moments of its spin-$\frac{3}{2}$ heavy quark spin symmetry counterparts, up to the next-to-leading order in covariant baryon chiral perturbation theory (BChPT) with the extended-on-mass-shell renormalization (EOMS) scheme. With the tree-level contributions fixed by the quark model while the two low energy constants (LECs) $C$ and $H$ controlling the loop contributions determined in two ways: the quark model (case 1) and lattice QCD simulations together with the quark model (case 2), we study the quark mass dependence of the magnetic moments and compare them with the predictions of the heavy baryon chiral perturbation theory (HB ChPT). It is shown that the difference is sizable in case 1, but not in case 2 due to the smaller LECs $C$ and $H$, similar to the case of spin-$\frac{1}{2}$ doubly charmed baryons. Second, we predict the magnetic moments of the spin-$\frac{3}{2}$ doubly charmed baryons and compare them with those of other approaches. The predicted magnetic moments in case 2 for the spin-$\frac{3}{2}$ doubly charmed baryons are closer to those of  other approaches. In addition, the large differences in case 1 and case 2 for the predicted magnetic moments may indicate the inconsistency between the quark model and the lattice QCD simulations, which should be checked by future experimental  or more lattice QCD data.
\end{abstract}

%\pacs{13.60.Le, 12.39.Mk,13.25.Jx}

\maketitle
\section{Introduction \label{sec:intro}}
One of the spin-$\frac{1}{2}$ doubly charmed baryons, $\Xi_{cc}^{+}$, was first reported by the SELEX Collaboration~\cite{Mattson:2002vu}. However, the following studies performed by the FOCUS~\cite{Ratti:2003ez}, Belle~\cite{Chistov:2006zj}, and BaBar~\cite{Aubert:2006qw} Collaborations found no evidence on its existence. In 2017, the LHCb Collaboration observed another spin-$\frac{1}{2}$ doubly charmed baryon $\Xi_{cc}^{++}$ in the decay mode $\Xi_{cc}^{++}\to\Lambda_c^+K^-\pi^+\pi^+$~\cite{Aaij:2017ueg}. The latest and more accurate measurement of its mass is $m_{\Xi_{cc}^{++}}=3621.55\pm0.23\pm0.30$~MeV~\cite{Aaij:2019uaz}. Since then, the properties, decay and production mechanisms of doubly charmed  baryons have been extensively studied theoretically~(see Refs.~\cite{Chen:2016spr,Liu:2019zoy} and references cited therein).

The masses of the $\frac{3}{2}$ doubly charmed baryons and their spin-$\frac{1}{2}$ counterparts are related to the $D^*$ and $D$ masses via the so-called heavy anti-quark
di-quark symmetry (HADS)~\cite{Hu:2005gf}, i.e.,
\begin{eqnarray}
&&m_{\Xi_{cc}^*}-m_{\Xi_{cc}}=\frac{3}{4}(m_{D^*}-m_D),\\
&&m_{\Omega_{cc}^*}-m_{\Omega_{cc}}=\frac{3}{4}(m_{D_s^*}-m_{D_s}).
\end{eqnarray}
These relations seem to be broken at the level of 25\% based on various theoretical and lattice QCD studies~\cite{Padmanath:2015jea,Chen:2017kxr,Alexandrou:2017xwd,Mathur:2018rwu}. Future discovery of the doubly charmed spin-$\frac{3}{2}$  baryons by the LHCb and Belle II experiments~\cite{Cerri:2018ypt,Kou:2018nap} will contribute tremendously to our understanding of the heavy quark symmetries as well as the non perturbative strong interaction.

The magnetic moments of baryons play a vital role in understanding their internal structure. For the spin-$\frac{1}{2}$ doubly charmed baryons, their magnetic moments have been systematically investigated in the heavy-baryon (HB) chiral perturbation theory (ChPT)~\cite{Li:2017cfz,Li:2020uok}, the extended on-mass shell (EOMS) BChPT~\cite{Liu:2018euh,Blin:2018pmj}, and the light-cone QCD sum rule (LCSR)~\cite{Ozdem:2018uue} after the discovery of $\Xi_{cc}^{++}$.
Up to now, the magnetic moments of spin-$\frac{3}{2}$ doubly charmed baryons have also been examined in a variety of phenomenological models~\cite{Lichtenberg:1976fi,Bose:1980vy,Bernotas:2012nz,Albertus:2006ya,Patel:2007gx,Dhir:2009ax,Sharma:2010vv}, the LCSR~\cite{Ozdem:2019zis}, the HB ChPT~\cite{Meng:2017dni}, and lattice QCD simulations~\cite{Can:2015exa}. It should be stressed that the lattice QCD study~\cite{Can:2015exa} only calculated the magnetic moment of $\Omega_{cc}^{*+}$ for an unphysical  $m_\pi\approx156~{\rm MeV}$. 

Compared to other phenomenological models, ChPT~\cite{Weinberg:1978kz,Gasser:1984gg,Gasser:1987rb,Scherer:2002tk} provides a systematic expansion of physical observables (magnetic moments in the present case) order by order. The chiral order ${n_\chi}$ is defined as $n_\chi=4L-2N_M-N_B+\sum_kkV_k$ for a given Feynman diagram with $L$ loops, $N_M~(N_B)$ internal meson~(baryon) propagators, and $V_k$ vertices from $k^{\rm th}$ order Lagrangians. In the one-baryon sector, because of the large nonzero baryon masses in the chiral limit, one needs to develop a power counting scheme different from that used for the mesonic sector. Over the years, three approaches have been developed and extensively studied, i.e., the HB~\cite{Jenkins:1990jv,Bernard:1995dp}, the infrared (IR)~\cite{Becher:1999he}, and the EOMS~\cite{Fuchs:2003qc} schemes. A brief summary and comparison of these three approaches can be found in Ref.~\cite{Geng:2013xn}.

The EOMS scheme has been successfully applied to study the magnetic moments of baryons~\cite{Geng:2009ys,Geng:2008mf,Geng:2009hh,Xiao:2018rvd,Liu:2018euh,Xiabng:2018qsd,Blin:2018pmj} in the past two decades, and it was shown that a better description of lattice QCD quark-mass dependent data can be achieved compared to the other two alternatives. In Refs.~\cite{Liu:2018euh,Blin:2018pmj}, the magnetic moments of the spin-$\frac{1}{2}$ doubly charmed baryons were studied in the EOMS BChPT. It was shown that the lattice QCD data of Ref.~\cite{Can:2013tna} can be described quite well. On the other hand, the extrapolated magnetic moments at the physical point are quite different from some of the phenomenological model predictions, which remain to be tested by future experimental measurements.

In the present work, we study the magnetic moments of the spin-$\frac{3}{2}$ doubly charmed baryons up to the next-to-leading order (NLO) in covariant baryon chiral perturbation theory (BChPT) with the extended-on-mass shell (EOMS) renormalization scheme. In addition to predicting the physical magnetic moments, we study their light-quark mass dependence such that they can be used to extrapolate future lattice QCD simulations to the physical point. The tree-level leading order contributions will be estimated by the quark model due to lack of experimental or lattice QCD data. The two low energy constants (LECs) $C$ and $H$ controlling the loop contributions are determined in two different ways: the quark model and the lattice QCD simulation supplemented by the quark model. In addition, to gain more insights into heavy quark spin symmetry, we compare the magnetic moments of the spin-$\frac{3}{2}$ doubly charmed baryons with those of their spin-$\frac{1}{2}$ counterparts, particularly, their light-quark mass dependence and the loop contributions via virtual spin-$\frac{3}{2}$ and -$\frac{1}{2}$ baryons.

This work is organized as follows. In Sec.~II, we introduce the electromagnetic form factors of the spin-$\frac{3}{2}$ baryons, provide the effective Lagrangians, and calculate the pertinent Feynman diagrams up to the next-to-leading order. In Sec.~III, we fix the relevant low energy constants with the help of the quark model and the lattice QCD simulation, predict the light-quark mass dependence of the magnetic moments, compare with previous studies, and examine the loop contributions. A short summary is given in Sec. IV.

\section{Theoretical formalism}
\label{sec:TH}
The matrix elements of the electromagnetic current $J_\mu$ for a spin-$\frac{3}{2}$ doubly charmed baryon can be parameterized as follows~\cite{Nozawa:1990gt,Geng:2009ys}:
\begin{eqnarray}
\langle T(p_f)|J_\mu|T(p_i)\rangle=&&-\bar{u}_\alpha(p_f)\left\{\left[\gamma_\mu F_1(\tau)+\frac{i\sigma_{\mu\nu}q^\nu}{2m_T}F_2(\tau)\right]g^{\alpha\beta}\right.\nonumber\\
&&\left.+\left[\gamma_\mu F_3(\tau)+\frac{i\sigma_{\mu\nu}q^\nu}{2m_T}F_4(\tau)\right]\frac{q^\alpha q^\beta}{4m_T^2}\right\}u_\beta(p_i),
\end{eqnarray}
where $\bar{u}_\alpha(p_f)$ and $u_\beta(p_i)$ are the Rarita-Schwinger~(RS) spinors~\cite{Rarita:1941mf}, $m_T$ is the doubly charmed baryon mass, and $\tau=-\frac{q^2}{4m_T^2}$. The four-momentum transfer is defined as $q=p_f-p_i$. The electric monopole~$G_{E0}$, quadrupole~$G_{E2}$, magnetic
dipole~$G_{M1}$, and octupole~$G_{M3}$ form factors can be expressed  in terms of the four electromagnetic form factors $F_i's~(i=1,2,3,4)$, respectively,
\begin{eqnarray}
&&G_{E0}(\tau)=\left[F_1(\tau)-\tau F_2(\tau)\right]+\frac{2}{3}\tau G_{E2}(\tau),\nonumber\\
&&G_{E2}(\tau)=\left[F_1(\tau)-\tau F_2(\tau)\right]-\frac{1}{2}(1+\tau)\left[F_3(\tau)-\tau F_4(\tau)\right],\nonumber\\
&&G_{M1}(\tau)=\left[F_1(\tau)+F_2(\tau)\right]+\frac{4}{5}\tau G_{M3}(\tau),\nonumber\\
&&G_{M3}(\tau)=\left[F_1(\tau)+F_2(\tau)\right]-\frac{1}{2}(1+\tau)\left[F_3(\tau)+F_4(\tau)\right].
\end{eqnarray}
At $q^2=0$, $Q_T=eG_{E0}(0)=eF_1(0)$ is the charge of the doubly charmed baryon, $\kappa_T=\frac{e}{2m_T}G_{M1}(0)=\frac{e}{2m_T}F_2(0)$ is the so-called anomalous magnetic moment, and the magnetic moment is defined as $\mu_T=\frac{m_N}{m_T}(\kappa_T+Q_T)$, where $m_N=940$ MeV is the nucleon mass. For the convenience of comparison with the magnetic moments of the doubly charmed baryons $\mu_T$ obtained in other approaches~\cite{Lichtenberg:1976fi,Bose:1980vy,Bernotas:2012nz,Oh:1991ws,Albertus:2006ya,Patel:2007gx,Dhir:2009ax,Sharma:2010vv,
Meng:2017dni,Ozdem:2019zis}, we take the nuclear magneton $\mu_N$ as the units of $\mu_T$ in this work.
\begin{figure}[h!]
  \centering
  \includegraphics[width=5cm]{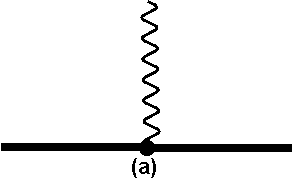}\\
  \includegraphics[width=5cm]{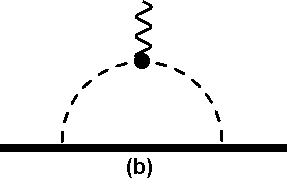}~~~\includegraphics[width=5.1cm]{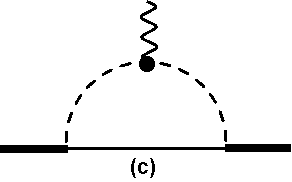}\\
  \includegraphics[width=5cm]{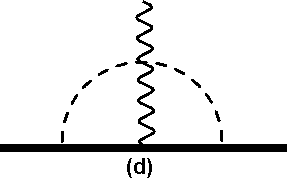}~~~\includegraphics[width=5.1cm]{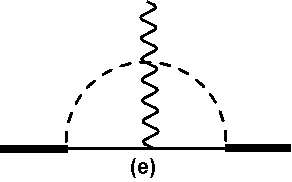}
  \caption{Feynman diagrams contributing to the spin-$\frac{3}{2}$ doubly charmed baryon magnetic moments up to NLO. Diagram (a) contributes at LO, diagrams (b)-(e) with a photon coupling to an intermediate meson or baryon contribute at NLO. The heavy~(light) solid, dashed, and wiggly lines denote spin-$\frac{3}{2}(\frac{1}{2})$ doubly charmed baryons, Goldstone bosons, and photons, respectively. The heavy dots represent the ${\cal O}(p^2)$ vertices.}\label{Figdiagram}
\end{figure}

In Fig.~\ref{Figdiagram}, diagrams~(a) and (b)-(e) contribute at ${\cal O}(p^2)$ and ${\cal O}(p^3)$, respectively. The leading order contribution to the magnetic moments $\mu_T$ is provided by the following Lagrangian:
\begin{eqnarray}
{\cal L}_{TT}^{(2)}=\frac{-ib_1^{tt}}{2m_T}\bar{T}^\mu\hat{F}_{\mu\nu}^+T^\nu+\frac{-ib_2^{tt}}{2m_T}\bar{T}^\mu T^\nu\mbox{\rm Tr}(F_{\mu\nu}^+),\label{eq:LagTree}
\end{eqnarray}
where the superscript in the Lagrangian stands for the chiral order, $b_1^{tt}$ and $b_2^{tt}$ are LECs, $\hat{F}_{\mu\nu}^+=F_{\mu\nu}^+-\frac{1}{3}{\rm Tr}(F_{\mu\nu}^+)$, $F_{\mu\nu}^+=|e|(u^\dag Q_HF_{\mu\nu}u+uQ_HF_{\mu\nu}u^\dag)$, $F_{\mu\nu}=\partial_\mu A_\nu-\partial_\nu A_\mu$,  $Q_H={\rm diag}(2,1,1)$ is the charge operator of the doubly charmed baryon, $u={\rm exp}[i\Phi/2F_\phi]$ with the unimodular matrix containing the pseudoscalar
nonet $\Phi$, and $F_\phi$ is the pseudoscalar meson decay constant. In principle, one can use either the chiral limit value or the physical value for the decay constant. In the present work, similar to Ref.~\cite{Meng:2017dni}, we choose to use the physical values, which are different for $\pi$, $K$, and $\eta$, i.e., $F_\pi=92.4~{\rm MeV}$, $F_K=1.22F_\pi$, and $F_\eta=1.3F_\pi$. In the present work, we denote the spin-$\frac{3}{2}$ and -$\frac{1}{2}$ doubly charmed baryons by $T^\mu$ and $B$, respectively,
\begin{eqnarray}
T^\mu=\left(
\begin{array}{c}
\Xi_{cc}^{*++}\\
\Xi_{cc}^{*+}\\
\Omega_{cc}^{*+}
\end{array}
\right),\qquad B=\left(
\begin{array}{c}
\Xi_{cc}^{++}\\
\Xi_{cc}^{+}\\
\Omega_{cc}^{+}
\end{array}
\right).
\end{eqnarray}

The loop diagrams contributing at NLO are determined by the lowest order Lagrangians ${\cal L}_B^{(1)}$+${\cal L}_T^{(1)}$+${\cal L}_{TB}^{(1)}$+${\cal L}_{BB}^{(1)}$+${\cal L}_M^{(2)}$, which are,
\begin{eqnarray}
{\cal L}_B^{(1)}&=&\bar{B}(i\slashed{D}-m_B)B,\nonumber\\
{\cal L}_T^{(1)}&=&\bar{T}^\mu\left[-g_{\mu\nu}(i\slashed{D}-m_T)+i(\gamma_\mu D_\nu+\gamma_\nu D_\mu)-\gamma_\mu(i\slashed{D}+m_T)\gamma_\nu\right] T^\nu,\nonumber\\
{\cal L}_{TB}^{(1)}&=&\frac{iC}{2m_TF_\phi}\left(\partial^\alpha\bar{T}^\mu\right)\gamma_{\alpha\mu\nu}B\partial^\nu\Phi+\mbox{H.c.},\nonumber\\
{\cal L}_{TT}^{(1)}&=&\frac{iH}{2m_TF_\phi}\bar{T}^\mu\gamma_{\mu\nu\rho\sigma}\gamma_5\left(\partial^\rho T^\nu\right)\partial^\sigma\Phi,\nonumber\\
{\cal L}_M^{(2)}&=&\frac{F_\phi^2}{4}{\rm Tr}[\nabla_\mu U(\nabla^\mu U)^\dag],
\end{eqnarray}
with
\begin{eqnarray}
&&D_\mu B=\partial_\mu B+\Gamma_\mu,\nonumber\\
&&\Gamma_\mu=\frac{1}{2}(u^\dag\partial_\mu u+u\partial_\mu u^\dag)-\frac{i}{2}(u^\dag v_\mu u+uv_\mu u^\dag)=-ieQ_HA_\mu,\nonumber\\
&&u_\mu=i(u^\dag\partial_\mu u-u\partial_\mu u^\dag)+(u^\dag v_\mu u-uv_\nu u^\dag),\nonumber\\
&&U=u^2=e^{\frac{i\Phi}{F_\phi}},\qquad\nabla_\mu U=\partial_\mu U+ieA_\mu[Q_l,U],
\end{eqnarray}
where $\gamma^{\mu\nu\alpha\beta}=\frac{1}{2}\left[\gamma^{\mu\nu\alpha},\gamma^\beta\right]$, $\gamma^{\mu\nu\alpha}=\frac{1}{2}\left\{\gamma^{\mu\nu},\gamma^{\alpha}\right\}$, $\gamma^{\mu\nu}=\frac{1}{2}\left[\gamma^\mu,\gamma^\nu\right]$, $v_\mu$ stands for the vector source, and the charge matrix for the light $u,d,s$ quarks is $Q_l={\rm diag}(2/3,-1/3,-1/3)$. Note that for the Lagrangians of $\bar{T}B\Phi$ and $\bar{T}T\Phi$, we use the ``consistent'' coupling scheme introduced in Refs.~\cite{ Pascalutsa:2006up,Pascalutsa:1999zz,Pascalutsa:2000kd}. It has been applied to study the magnetic moments of octet baryons~\cite{Geng:2009hh}, decuplet baryons~\cite{Geng:2009ys}, and singly charmed baryons~\cite{Xiabng:2018qsd}, which can provide a proper description of the experimental/lattice QCD data and converges relatively faster.

The leading order tree-level contributions to the magnetic moments of the doubly charmed baryons $T^\mu$ can be obtained from Eq.~(\ref{eq:LagTree}) as follows:
\begin{eqnarray}
\kappa_T^{(2)}=\alpha b_1^{tt}+\beta b_2^{tt},\label{TreeMM}
\end{eqnarray}
with the values of $\alpha$ and $\beta$ listed in Table~\ref{eq:TreeCCs}.
\begin{table}[h!]
 \caption{\label{eq:TreeCCs}Coefficients of the tree-level contributions in Eq.~(\ref{TreeMM}).}
\begin{center}
    \begin{tabular}{cccc}
      \hline
      \hline
      ~~~~~~ & ~~~~~~$\Xi_{cc}^{*++}$~~~~~~ & ~~~~~~$\Xi_{cc}^{*+}$~~~~~~ & ~~~~~~$\Omega_{cc}^{*+}$~~~~~~\\
      \hline
      $\alpha$ & $\frac{4}{3}$ & $-\frac{2}{3}$ & $-\frac{2}{3}$\\
      \hline
      $\beta$ & $8$ & $8$ & $8$\\
      \hline
      \hline
    \end{tabular}
  \end{center}
\end{table}

The loop diagrams of Figs.~\ref{Figdiagram} (b), (c), (d), and (e) contribute to the magnetic moments at ${\cal O}(p^3)$, which are written as,
\begin{eqnarray}
\kappa_T^{(3)}&=&\sum_{\phi=\pi,K}\frac{H^2}{F_\phi^2}\xi_{T\phi}^{(3,b)}H_T^{(b)}(m_\phi)
+\sum_{\phi=\pi,K}\frac{C^2}{F_\phi^2}\xi_{T\phi,\delta}^{(3,c)}H_T^{(c)}(\delta,m_\phi)\nonumber\\
&&+\sum_{\phi=\pi,K,\eta}\frac{H^2}{F_\phi^2}\xi_{T\phi}^{(3,d)}H_T^{(d)}(m_\phi)
+\sum_{\phi=\pi,K,\eta}\frac{C^2}{F_\phi^2}\xi_{T\phi,\delta}^{(3,e)}H_T^{(e)}(\delta,m_\phi),\label{LoopMM}
\end{eqnarray}
with the coefficients $\xi_{T\phi,\delta_i}^{(3;b,c,d,e)}$ tabulated in Table~\ref{tab:LoopCCs}. Here, $\delta=m_T-m_B$ is the mass difference between the spin-$\frac{3}{2}$ and spin-$\frac{1}{2}$ doubly charmed baryons. The loop functions $H_T^{(b,d)}(m_\phi)$ and $H_T^{(c,e)}(\delta,m_\phi)$ correspond to $\frac{16\pi^2}{m_T^2}\left(H_2^{(g,h)}-H_{\rm PC}^{(g,h)}\right)$ and $\frac{16\pi^2}{m_T^2}\left(H_2^{(d,e)}-H_{\rm PC}^{(d,e)}\right)$, which can be found in the Appendix of Ref.~\cite{Geng:2009ys}. The power counting breaking~(PCB) terms $H_{\rm PC}^{(g,h)}$ and $H_{\rm PC}^{(d,e)}$ are determined by expanding $m_\phi$ up to ${\cal O}(p^0)$. Note that the pertinent loop functions are regularized with the $\widetilde{MS}$ scheme.

\begin{table}[h!]
 \caption{\label{tab:LoopCCs}Coefficients of the loop contributions in Eq.~(\ref{LoopMM}) for the $T^\mu$ states.}
\begin{center}
    \begin{tabular}{cccc}
      \hline
      \hline
      ~~~~~~ & ~~~~~~$\Xi_{cc}^{*++}$~~~~~~ & ~~~~~~$\Xi_{cc}^{*+}$~~~~~~ & ~~~~~~$\Omega_{cc}^{*+}$~~~~~~\\
      \hline
      $\xi_{T\pi}^{(3,b)}$ & $\frac{1}{2}$ & $-\frac{1}{2}$ & $0$\\
      \hline
      $\xi_{TK}^{(3,b)}$ & $\frac{1}{2}$ & $0$ & $-\frac{1}{2}$\\
      \hline
      $\xi_{T\pi,\delta}^{(3,c)}$ & $\frac{1}{2}$ & $-\frac{1}{2}$ & $0$\\
      \hline
      $\xi_{TK,\delta}^{(3,c)}$ & $\frac{1}{2}$ & $0$ & $-\frac{1}{2}$\\
      \hline
      $\xi_{T\pi}^{(3,d)}$ & $1$ & $\frac{5}{4}$ & $0$\\
      \hline
      $\xi_{TK}^{(3,d)}$ & $\frac{1}{2}$ & $\frac{1}{2}$ & $\frac{3}{2}$\\
      \hline
      $\xi_{T\eta}^{(3,d)}$ & $\frac{1}{6}$ & $\frac{1}{12}$ & $\frac{1}{3}$\\
       \hline
      $\xi_{T\pi,\delta}^{(3,e)}$ & $1$ & $\frac{5}{4}$ & $0$\\
      \hline
      $\xi_{TK,\delta}^{(3,e)}$ & $\frac{1}{2}$ & $\frac{1}{2}$ & $\frac{3}{2}$\\
      \hline
      $\xi_{T\eta,\delta}^{(3,e)}$ & $\frac{1}{6}$ & $\frac{1}{12}$ & $\frac{1}{3}$\\
      \hline
      \hline
    \end{tabular}
  \end{center}
\end{table}

In addition, we can obtain the HB counterparts of the loop functions  by performing $1/m_T$ expansions for the loop functions obtained in the EOMS scheme, which turn out to agree with those of Ref.~\cite{Meng:2017dni}.

\section{Results and discussions}
\label{sec:Results}
In the following numerical analysis, we take the masses of the spin-$\frac{1}{2}$ doubly charmed baryons to be $m_B=m_{\Xi_{cc}^{+}}=3.62~{\rm GeV}$~\cite{Zyla:2020zbs}. Here, we do not consider the mass difference among the doubly charmed baryon triplet, in other words, we neglect SU(3) symmetry breakings in the masses, which are of higher chiral order. On the other hand, although the spin-$\frac{3}{2}$ doubly charmed baryons have not been observed, many theoretical studies~\cite{Bagan:1992za,Roncaglia:1995az,SilvestreBrac:1996bg,Ebert:1996ec,Tong:1999qs,Gershtein:2000nx,Kiselev:2001fw,Kiselev:2002iy,
Narodetskii:2001bq,Ebert:2002ig,Mathur:2002ce,Albertus:2006ya,Martynenko:2007je,Tang:2011fv,Roberts:2007ni,Valcarce:2008dr,Karliner:2014gca,
Shah:2016vmd,Chen:2016spr,Hu:2005gf,Lu:2017meb,Xiao:2017udy,Patel:2007gx,Alexandrou:2012xk,Namekawa:2013vu,Kiselev:2017eic,Migura:2006ep,
Chiu:2005zc,Vijande:2004at,Flynn:2003vz,Lewis:2001iz,Itoh:2000um,Bahtiyar:2018vub,Bahtiyar:2020uuj} have shown that the mass splitting $\delta$ between the spin-$\frac{3}{2}$ and spin-$\frac{1}{2}$ doubly charmed baryons is in the range of  a few dozens MeV to 100 MeV. In this work, we choose $\delta=100~{\rm MeV}$. Nevertheless, we have performed calculations with different values for $\delta$  and found that the magnetic moments are not very sensitive to $\delta$, consistent with the study of Ref.~\cite{Meng:2017dni}. The mass difference $\delta$ vanishes in the heavy quark mass limit.  For the masses of the pseudoscalar mesons, we use the PDG values~\cite{Zyla:2020zbs}. In addition, we fix the renormalization scale at $\mu=1$~GeV. We have checked that the magnetic moments in both the EOMS BChPT and HB ChPT are almost independent of $\mu$.

We do not have  experimental or adequate lattice QCD data to determine the LECs $b_1^{tt}$ and $b_2^{tt}$. As a result, we turn to the quark model, where the magnetic moments can be calculated from the sum of the magnetic moments of the three constituent quarks, i.e.,
\begin{eqnarray}
\mu_q=\frac{e_q}{2m_q}=Q_q\frac{m_N}{m_q},\qquad (q=u,d,s,c)\label{quarkmodelMM}
\end{eqnarray}
where $\mu_q$ is in units of the nuclear magneton $\mu_N$, $Q_q$ stands for the quark charge, and $m_q$ is the constituent quark mass. In this work, we take the constituent quark masses from Ref.~\cite{Lichtenberg:1976fi}, which are $m_u=m_d=336$~MeV, $m_s=540$~MeV, and $m_c=1660$~MeV~\footnote{In Ref.~\cite{Meng:2017dni}, the authors estimated the LO magnetic moments of the spin-$\frac{3}{2}$ doubly charmed baryons using two sets of constituent quark masses~\cite{Lichtenberg:1976fi,Karliner:2014gca} and found that the so-obtained results are consistent with each other.}. Using Eq.~(\ref{quarkmodelMM}), one can easily obtain the LO contributions to the magnetic moments of the spin-$\frac{3}{2}$ doubly charmed baryons as follows:
\begin{eqnarray}
&&\mu_{\Xi_{cc}^{*++}}^{(2)}=\frac{m_N}{m_{\Xi_{cc}^{*++}}}(2+\kappa_{\Xi_{cc}^{*++}}^{(2)})=2\mu_c+\mu_u=2.61,\nonumber\\
&&\mu_{\Xi_{cc}^{*+}}^{(2)}=\frac{m_N}{m_{\Xi_{cc}^{*+}}}(1+\kappa_{\Xi_{cc}^{*+}}^{(2)})=2\mu_c+\mu_d=-0.18,\nonumber\\
&&\mu_{\Omega_{cc}^{*+}}^{(2)}=\frac{m_N}{m_{\Omega_{cc}^{*+}}}(1+\kappa_{\Omega_{cc}^{*+}}^{(2)})=2\mu_c+\mu_s=0.17.\label{LOMMnum}
\end{eqnarray}
Note that, the quark model predictions break the SU(3) flavor symmetry, because of the use of different quark masses. While in the leading order chiral Lagrangian of Eq.~(\ref{eq:LagTree}) which respects SU(3) symmetry, only two LECs are available. As a result, we could not fix the two leading order LECs by the quark model. In this work, we simply use the quark model predictions as the leading order results for the EOMS BChPT.
The results in Eq.~(\ref{LOMMnum}) show that the LO magnetic moments of $\Xi_{cc}^{*+}$ and $\Omega_{cc}^{*+}$ are relatively small. The reason for this is that the contributions of one light and two heavy quarks cancel each other due to the opposite quark charge. The three quark charges in $\Xi_{cc}^{*++}$ are all positive, and therefore the LO magnetic moment of $\Xi_{cc}^{*++}$ is the largest. This point has also been  noted in Ref.~\cite{Meng:2017dni}.

For the contributions of the loop diagrams, we determine the relevant LECs $C$ and $H$ in two ways. In the first case, the two LECs are related to the axial-vector coupling of the nucleon $g_A$ with the help of the quark model~\cite{Li:2017cfz,Li:2017pxa}, i.e., $C=-\frac{2\sqrt{3}}{5}g_A$, $H=-\frac{3}{5}g_A$, and $g_A=1.267$. In the second scenario, we define a common factor $\rho$  to re-scale  the LECs $C$ and $H$ simultaneously. In this way, we have $C=-\frac{2\sqrt{3}}{5}g_A\cdot\rho$ and $H=-\frac{3}{5}g_A\cdot\rho$. The parameter $\rho$ can be determined by the lattice QCD result for the magnetic moment of $\Omega_{cc}^{*+}$~\cite{Can:2015exa} with the leading order contributions determined by  the quark model.

\subsection{The magnetic moments of spin-$\frac{3}{2}$ charmed baryons in case 1}
\begin{figure}[h!]
  \centering
  \includegraphics[width=10cm]{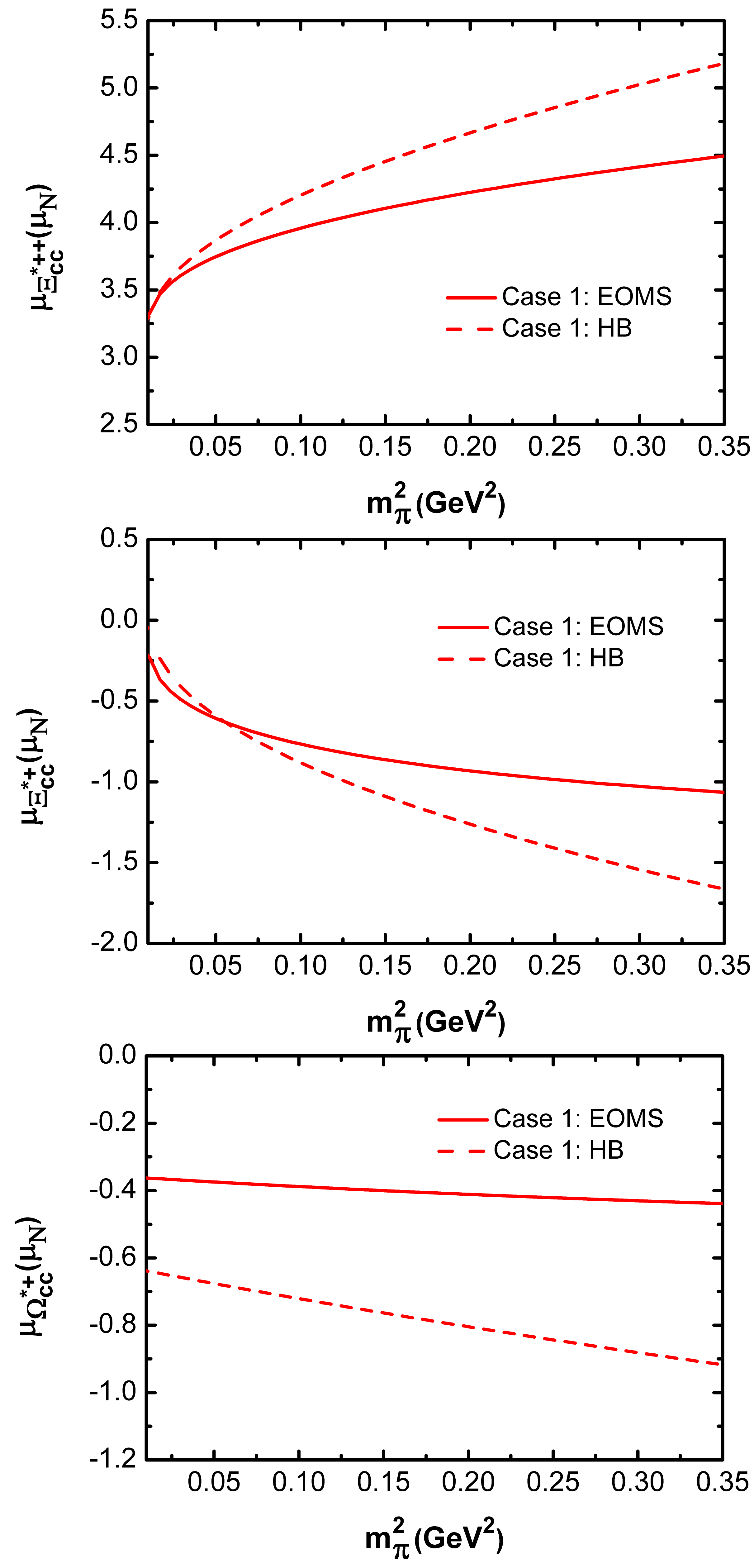}\\
  \caption{Magnetic moments of the spin-$\frac{3}{2}$ charmed baryons as a function of $m_\pi^2$ in case 1. The solid and dashed lines in red represent  the results of the EOMS BChPT and HB ChPT, respectively.}\label{FigMMpi2}
\end{figure}
In this case, the tree and loop level contribution to the magnetic moments of the spin-$\frac{3}{2}$ doubly charmed baryons are estimated by the quark model, as explained above. In Fig.~\ref{FigMMpi2}, we plot the magnetic moments of the spin-$\frac{3}{2}$ doubly charmed baryons $\mu_T$ as a function of $m_\pi^2$. It is shown that there is some distinct difference between the EOMS BChPT and HB ChPT results. In general, the light quark mass dependence is milder in the EOMS scheme than in the HB scheme. It should be noted that in Ref.~\cite{Liu:2018euh} the HB and EOMS results are almost identical. This is because  for the spin-$\frac{1}{2}$ doubly charmed baryons, the loop contributions are much suppressed because of the small coupling $g_a=0.08$ determined by the lattice QCD data, $g_a=0.2$ by the heavy antiquark diquark symmetry, and $g_a=0.25$ by the quark model~\cite{Liu:2018euh}. For the spin-$\frac{3}{2}$ doubly charmed baryons, the two LECs $H=-0.76$ and $C=-0.88$  are more than three times larger, resulting in a loop contribution of more than 10 times larger. Therefore, the relatively large LECs $H$ and $C$ are responsible for the large difference between the EOMS BChPT and HB ChPT results. The difference for the $\Omega_{cc}^{*+}$ baryon is the most notable. 
\begin{table}[h!]
 \caption{\label{tab:loopDecomposition}Decomposition of the loop contributions to the magnetic moments of spin-$\frac{3}{2}$ doubly charmed baryons in case 1. The subscript $T$ and $B$ denote the loop diagrams with intermediate $T^\mu$ and $B$ states at ${\cal O}(p^3)$, respectively.}
\begin{center}
    \begin{tabular}{ccccccccccc}
      \hline
      \hline
      \multirow{2}{0.5cm}  &  & \multicolumn{4}{c}{Case 1: EOMS} &  & \multicolumn{4}{c}{Case 1: HB}\\
      \cline{3-6}\cline{8-11}
      &  & ~~~~${\cal O}(p^2)$~~~~ & ~~~~${\cal O}(p^3)_T$~~~~ & ~~~~${\cal O}(p^3)_B$~~~~ & ~~~~$\mu_{\rm tot}$~~~~ & & ~~~~${\cal O}(p^2)$~~~~ & ~~~~${\cal O}(p^3)_T$~~~~ & ~~~~${\cal O}(p^3)_B$~~~~ & ~~~~$\mu_{\rm tot}$~~~~\\
      \hline
      \multicolumn{2}{c}{$\mu_{\Xi_{cc}^{*++}}$} & $2.61$ & $0.27$ & $0.62$ & $3.50$ &  & $2.61$ & $0.39$ &$0.51$ & $3.51$\\
      \hline
      \multicolumn{2}{c}{$\mu_{\Xi_{cc}^{*+}}$} & $-0.18$ & $-0.08$ & $-0.13$ & $-0.39$ &  & $-0.18$ & $-0.11$ & $0.02$ & $-0.27$\\
      \hline
      \multicolumn{2}{c}{$\mu_{\Omega_{cc}^{*+}}$} & $0.17$ & $-0.15$ & $-0.39$ & $-0.37$ &  & $0.17$ & $-0.27$ & $-0.54$ & $-0.64$\\
      \hline
      \hline
    \end{tabular}
  \end{center}
\end{table}

In Table~\ref{tab:loopDecomposition}, we decompose the loop contributions to the magnetic moments $\mu_T$ into those from the intermediate $T^\mu$ and $B$ states, respectively. For the spin-$\frac{3}{2}$ doubly charmed baryons, it should be noted that the absolute contributions of the intermediate $B$ baryons are more than those of intermediate $T^\mu$ baryons and their contributions are of the same sign in the EOMS BChPT and HB ChPT except for the $\Xi_{cc}^{*+}$ baryon in the HB ChPT. It indicates that loop corrections are important and non-negligible. Interestingly, for $\Omega_{cc}^{*+}$ the loop correction is much larger than the tree-level contribution, which can be attributed to the larger $H$ and $C$ predicted by the quark model, as also noted in Ref.~\cite{Meng:2017dni}. It should be  mentioned that in the study of the spin-$\frac{1}{2}$ doubly charmed baryons~\cite{Liu:2018euh} the authors found that the coupling $g_a$ determined by the lattice QCD data~\cite{Can:2013tna} is much smaller than the one predicted by the quark model. Similar thing could happen here, as shown below. In a recent work on spin-$\frac{1}{2}$ doubly charmed baryons in the HB ChPT~\cite{Li:2020uok}, it is shown that the loop contributions of intermediate $T^\mu$ and $B$ cancel each other, which is opposite to the case of spin-$\frac{3}{2}$ baryons. This points to the importance of loop corrections again for spin-$\frac{3}{2}$ states.

\subsection{The magnetic moment of spin-$\frac{3}{2}$ charmed baryons in case 2}
\begin{figure}[h!]
  \centering
  \includegraphics[width=10cm]{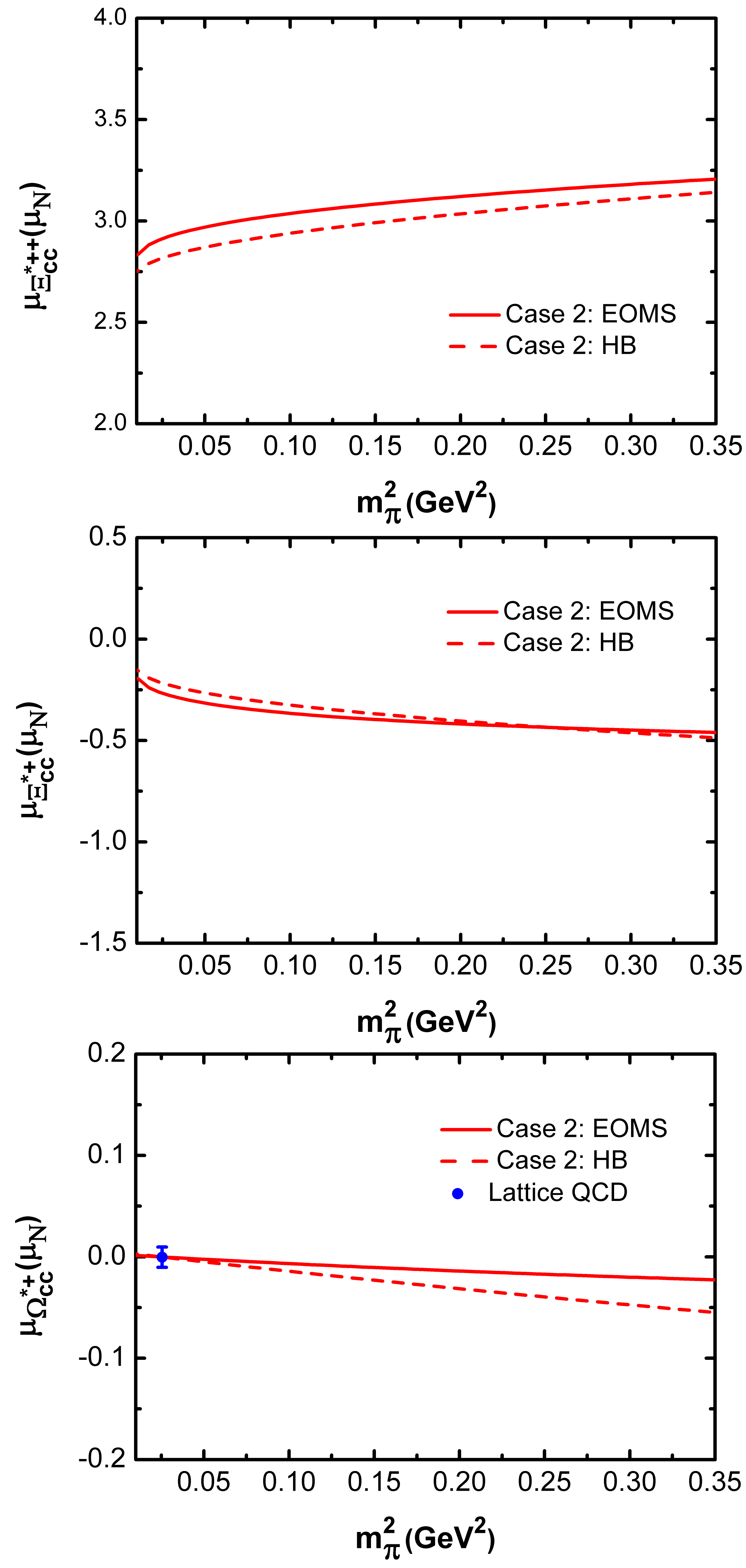}\\
  \caption{Magnetic moments of the spin-$\frac{3}{2}$ charmed baryons as a function of $m_\pi^2$ in case 2. The data point in blue stands for the lattice QCD unphysical value at $m_\pi\approx156~{\rm MeV}$. The solid and dashed lines in red represent the results of the EOMS BChPT and HB ChPT, respectively.}\label{FigMMpi2LQCD}
\end{figure}
In a lattice QCD simulation~\cite{Can:2015exa}, Can et al. studied the magnetic moment of $\Omega_{cc}^{*+}$ for a pion mass of $\approx156~{\rm MeV}$ and they obtained   $\mu_{\Omega_{cc}^{*+}}=0.000(10)$, which is rather different from the predictions of the HB ChPT and EOMS BChPT shown in the previous section. However, keeping in mind that the BChPT predictions rely on the quark model inputs, both at tree level and one-loop level. As there are four unknow LECs in the BChPT, we could not determine all of them using the lattice QCD magnetic moment for $\Omega_{cc}^{*+}$.  Therefore, in the present work, we  tentatively assume that the quark model predicted magnetic moments are resonable, which are taken as the leading order ChPT results, and use the only lattice QCD datum to fix the strength of the loop functions, i.e., $C$ and $H$. As the one lattice QCD magnetic moment can not fix two LECs, we further assume that they are re-scaled by a same factor, namely, we multiply both LECs $C$ and $H$ by a common factor $\rho$, leading to effectively only one unknown parameter $\rho$ which can be determined by reproducing the single lattice QCD magnetic moment for $\Omega_{cc}^{*+}$. This way,  we obtain $\rho_{\rm EOMS}=0.563(17)$ and $\rho_{\rm HB}=0.455(13)$ in the EOMS BChPT and HB ChPT, respectively. As a result, the LECs $C$ and $H$ are almost reduced by half compared to the predictions of the quark model. Using them, we replot the magnetic moments of the spin-$\frac{3}{2}$ doubly charmed baryons $\mu_T$ as a function of $m_\pi^2$. As shown in Fig.~\ref{FigMMpi2LQCD}, the HB and EOMS results are not much different, which is similar to the case of  spin-$\frac{1}{2}$ doubly charmed baryons~\cite{Liu:2018euh}. This indicates that the quark model predictions for $C$ and $H$ might be overestimated, which should be verified by future experimental or more lattice QCD data.
\begin{table}[h!]
 \caption{\label{tab:loopDecompositionLQCD}Decomposition of the loop contributions to the magnetic moments of spin-$\frac{3}{2}$ doubly charmed baryons in case 2. The subscript $T$ and $B$ denote the loop diagrams with intermediate $T^\mu$ and $B$ states at ${\cal O}(p^3)$, respectively.}
\begin{center}
    \begin{tabular}{ccccccccccc}
      \hline
      \hline
      \multirow{2}{0.5cm}  &  & \multicolumn{4}{c}{Case 2: EOMS} &  & \multicolumn{4}{c}{Case 2: HB}\\
      \cline{3-6}\cline{8-11}
      &  & ~~~~${\cal O}(p^2)$~~~~ & ~~~~${\cal O}(p^3)_T$~~~~ & ~~~~${\cal O}(p^3)_B$~~~~ & ~~~~$\mu_{\rm tot}$~~~~ & & ~~~~${\cal O}(p^2)$~~~~ & ~~~~${\cal O}(p^3)_T$~~~~ & ~~~~${\cal O}(p^3)_B$~~~~ & ~~~~$\mu_{\rm tot}$~~~~\\
      \hline
      \multicolumn{2}{c}{$\mu_{\Xi_{cc}^{*++}}$} & $2.61$ & $0.08$ & $0.20$ & $2.89$ &  & $2.61$ & $0.08$ &$0.11$ & $2.80$\\
      \hline
      \multicolumn{2}{c}{$\mu_{\Xi_{cc}^{*+}}$} & $-0.18$ & $-0.03$ & $-0.04$ & $-0.25$ &  & $-0.18$ & $-0.02$ & $0.00$ & $-0.20$\\
      \hline
      \multicolumn{2}{c}{$\mu_{\Omega_{cc}^{*+}}$} & $0.17$ & $-0.04$ & $-0.12$ & $0.001$ &  & $0.17$ & $-0.05$ & $-0.11$ & $0.001$\\
      \hline
      \hline
    \end{tabular}
  \end{center}
\end{table}

Next, we also decompose the loop contributions to the magnetic moments $\mu_T$. As can be seen in Table~\ref{tab:loopDecompositionLQCD}, the predictions of the magnetic moments for all the spin-$\frac{3}{2}$ states are smaller in absolute value than those obtained by the quark model in case 1 due to the reduced loop contributions. Especially, the puzzling feature that the loop correction for $\Omega_{cc}^{*+}$ is much larger than the tree-level contribution in case 1 has disappeared. Nevertheless, the contributions of the intermediate $B$ baryons and $T^\mu$ baryons still add coherently  in the EOMS BChPT and HB ChPT, indicating the importance of the loop contributions.
\begin{figure}[h!]
  \centering
  \includegraphics[width=10cm]{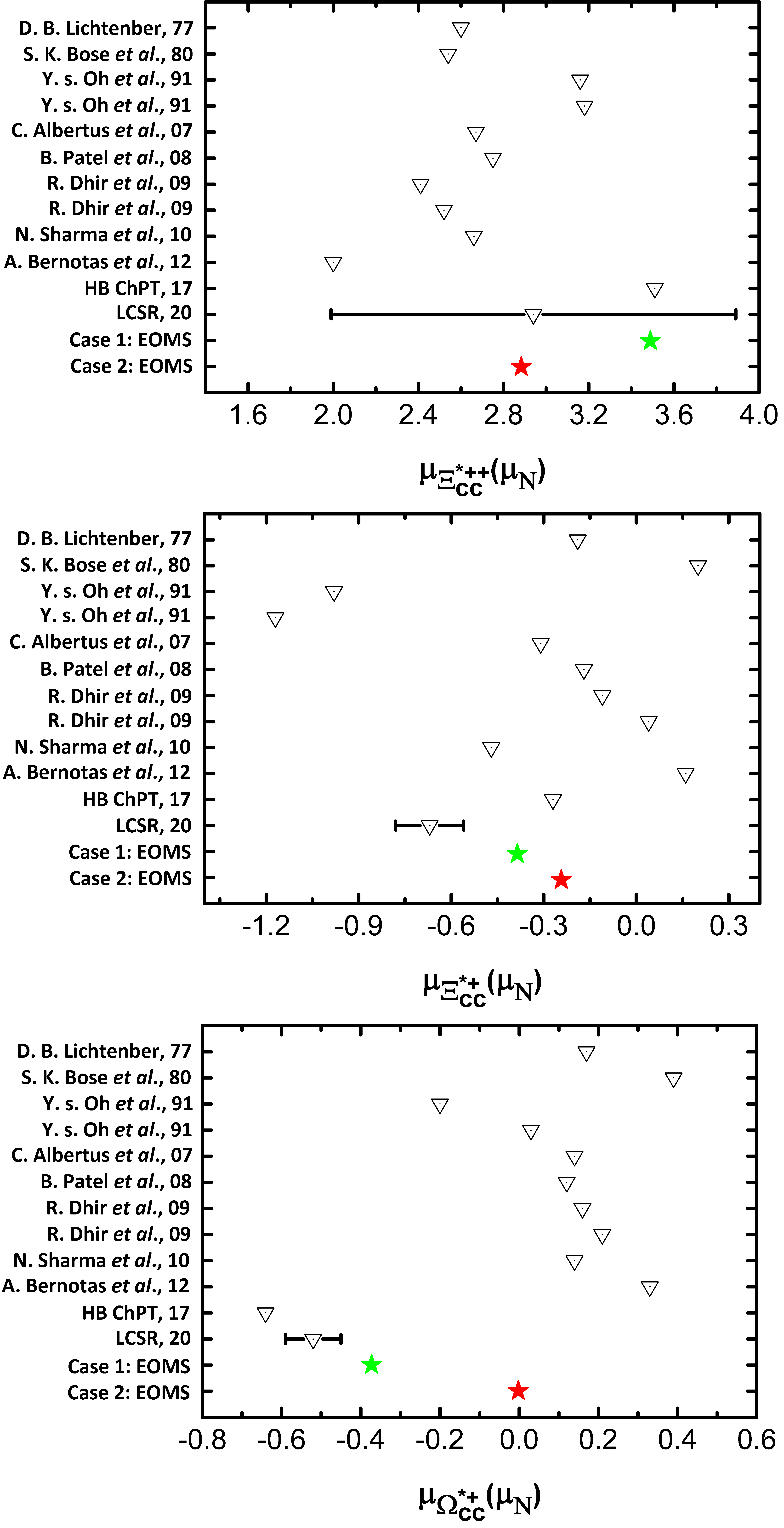}\\
  \caption{Magnetic moments of the spin-$\frac{3}{2}$ charmed baryons obtained in different approaches. The solid stars denote
   the results predicted in the EOMS BChPT in two different ways. The others are taken from the quark model~\cite{Lichtenberg:1976fi} (D. B. Lichtenber, 77), the bag model~\cite{Bose:1980vy,Bernotas:2012nz} (S. K. Bose et al., 80 and A. Bernotas et al., 12), the Skymion model~\cite{Oh:1991ws} (Y. s. Oh et al., 91), non-relativistic quark model~\cite{Albertus:2006ya} (C. Albertus et al., 07), hyper central model~\cite{Patel:2007gx} (B. Patel et al., 08), effective mass and screened charge scheme~\cite{Dhir:2009ax} (R. Dhir et al., 09), the chiral constituent quark model~\cite{Sharma:2010vv} (N. Sharma et al., 10), the HB ChPT~\cite{Meng:2017dni} (HB ChPT, 17), and the light-cone QCD sum rule (LCSR)~\cite{Ozdem:2019zis} (LCSR, 20).}\label{FigMMdiff}
\end{figure}

In Fig.~\ref{FigMMdiff} we compare our predicted magnetic moments with those obtained by other approaches. The predicted magnetic moments for the three spin-$\frac{3}{2}$ states using the EOMS BChPT in case 2 are: $\mu_{\Xi_{cc}^{*++}}=2.891(16)$, $\mu_{\Xi_{cc}^{*+}}=-0.248(4)$, and  $\mu_{\Omega_{cc}^{*+}}=0.001(10)$. The predictions in case 1 can be found in Table.~\ref{tab:loopDecomposition}. The uncertainties in case 2 originate from the factor $\rho_{\rm EOMS}$ determined by the lattice QCD data. One finds that the magnetic moments predicted in case 2 are closer to the results of other approaches. The large differences of magnetic moments predicted in case 1 and case 2 may indicate the inconsistency between the quark model and the lattice QCD simulations. Hopefully, future lattice QCD or experimental studies can clarify the interesting situation. 

\subsection{Heavy quark symmetry and its breaking}

Here, we discuss the heavy quark symmetry and its breaking and their impact on the magnetic moments of the $T^\mu$ baryons at ${\cal O}(p^3)$. In the heavy quark mass limit, i.e., $\delta=0$, the loop contributions of intermediate $B$ baryons are exactly twice as much as those of intermediate $T^\mu$ baryons. This can be checked by setting $\delta=0$ in the HB ChPT loop functions. Compared to the HB ChPT, loop functions in the EOMS BChPT include  relativistic corrections which break the heavy quark symmetry. For the diagrams with a photon attached to an intermediate meson, the loop contributions for $\delta=0$ can be expressed as:
\begin{eqnarray}
&&H^2\cdot H_T^{(b)}(m_\phi)=\frac{9}{25}g_A^2m_T\cdot\left(\frac{m_\pi}{12\pi}m_\phi-\frac{36\log\left(\frac{m_T^2}{m_\phi^2}\right)-49}{144\pi^2 m_T}m_\phi^2+{\cal O}\left(\frac{1}{m_T^2}\right)\right),\\
&&C^2\cdot H_T^{(c)}(0,m_\phi)=\frac{12}{25}g_A^2m_T\cdot\left(\frac{m_\pi}{8\pi}m_\phi-\frac{27\log\left(\frac{m_T^2}{m_\phi^2}\right)-28}{96\pi^2 m_T}m_\phi^2+{\cal O}\left(\frac{1}{m_T^2}\right)\right).
\end{eqnarray}
It can be clearly seen that the contributions of intermediate $B$ baryons are twice as much as those of intermediate $T^\mu$ baryons at the order of $\left(\frac{1}{m_T}\right)^0$. However, at higher orders $\left(\frac{1}{m_T}\right)^n~(n\geq1)$ the relation is broken. For the
diagrams with a photon attached to an intermediate baryon, one can obtain the contribution for $\delta=0$ as follows:
\begin{eqnarray}
&&H^2\cdot H_T^{(d)}(m_\phi)=\frac{9}{25}g_A^2m_T\cdot\left(-\frac{21\log\left(\frac{m_T^2}{m_\phi^2}\right)-3\log\left(\frac{m_\phi^2}{\mu^2}\right)-167}{864\pi^2 m_T}m_\phi^2+{\cal O}\left(\frac{1}{m_T^2}\right)\right),\nonumber\\
&&C^2\cdot H_T^{(e)}(0,m_\phi)=\frac{12}{25}g_A^2m_T\cdot\left(\frac{5}{48\pi^2m_T}m_\phi^2+{\cal O}\left(\frac{1}{m_T^2}\right)\right).\label{BpHQS}
\end{eqnarray}
From  Eq.~(\ref{BpHQS}), we find that the contributions start at $\frac{1}{m_T}$. These higher order relativistic corrections break the heavy quark symmetry relation as well. Note that the contribution of the diagram with a photon attached to an intermediate baryon vanishes at ${\cal O}(p^3)$ in the HB ChPT and the $m_T$ in the above equations outside the brackets is from the baryon field normalization.

\section{Summary}
We studied the magnetic moments of the spin-$\frac{3}{2}$ doubly charmed baryons in the covariant baryon chiral perturbation theory (BChPT) up to the next-to-leading order. To recover the power counting, we adopted the extended-on-mass-shell renormalization (EOMS) scheme. Due to lack of experimental and lattice QCD data, the leading order contributions to the magnetic moments are fixed by the quark model. The other two low energy constants $C$ and $H$ which appear in the loop contributions are determined by two ways: the quark model (case 1) and the lattice QCD data supplemented by the quark model (case 2).

To facilitate future lattice QCD simulations, we predicted the magnetic moments of the spin-$\frac{3}{2}$ doubly charmed baryons as a function of $m_\pi^2$. In case 1, the larger couplings $C$ and $H$~(compared to the case of the spin-$\frac{1}{2}$ doubly charmed baryons) resulted in large differences between the EOMS BChPT and HB ChPT results. The predicted magnetic moment for  $\Omega_{cc}^{*+}$ is found to differ from that predicted by the lattice QCD simulations for $m_\pi\approx156~{\rm MeV}$. The discrepancy can be reconciled by suppressing the loop contributions. We show that if the quark model predicted $H$ and $C$ is reduced by half, one can obtain a magnetic moment for $\Omega_{cc}^{*+}$ consistent with that of the lattice QCD. In this case, the quark mass dependencies of the magnetic moments are much smaller than those of case 1.  It should be noted that  our predicted magnetic moments in case 2 for the three spin-$\frac{3}{2}$ states are closer to those of the majority of other approaches.

An interesting discovery is that for the magnetic moments of the spin-$\frac{3}{2}$ doubly charmed baryons, the contributions at ${\cal O}(p^3)$ of intermediate $T^\mu$ and $B$ baryons add coherently, instead destructively, as for the spin-$\frac{1}{2}$ doubly charmed baryons. Meanwhile, in the heavy quark mass limit, the intermediate $B$ contribution is twice that of the $T^\mu$ contribution. In reality, this relation is broken by the finite heavy quark masses and relativistic corrections. The coherent interference  of the loop contributions of intermediate $T^\mu$ and $B$ baryons  indicates that the loop contributions are relatively more important, compared to the case of spin-$\frac{1}{2}$ doubly charmed baryons.

\section{Acknowledgements}
We thank Dr. Kadir Utku Can
for reminding us of Ref.~\cite{Can:2015exa}. R. X. S. thanks Lu Meng for useful discussions. This work was partly supported by the National Natural Science Foundation of China (NSFC) under Grants No. 11975041, No.11735003, No.11961141004, and the Academic Excellence  Foundation of BUAA for Ph.D. Students.

\bibliography{MMdoublyc}

\end{document}